\documentclass[journal]{IEEEtran}

\ifCLASSINFOpdf
\else
   \usepackage[dvips]{graphicx}
\fi
\usepackage{url}

\usepackage{amsmath,amssymb,amsfonts,graphicx}
\usepackage{cite}
\usepackage[ruled,norelsize]{algorithm2e}
\SetKwBlock{Repeat}{repeat}{}
\SetKwFor{At}{at}{do}{end}
\usepackage{array}
\usepackage[font=small,labelfont=bf]{caption}
\usepackage{subcaption}
\usepackage{dsfont}
\usepackage{comment}
\usepackage{hhline}
\usepackage[flushleft]{threeparttable}
\usepackage{nccmath}
\usepackage{multicol}
\usepackage{makecell}
\usepackage{tabularx}
\usepackage[hidelinks]{hyperref}
\usepackage{xcolor}
\usepackage{amsthm}

\SetKwFor{While}{while}{}{end while}%
\SetKwFor{For}{for}{}{end for}%

\newcolumntype{Y}{>{\centering\arraybackslash}X}
\newcolumntype{b}{>{\hsize=1.2\hsize}Y}
\newcolumntype{z}{>{\hsize=1.30\hsize}Y}
\newcolumntype{s}{>{\hsize=.5\hsize}Y}

\makeatletter
\newcommand{\removelatexerror}{\let\@latex@error\@gobble}
\makeatother

\renewcommand{\baselinestretch}{0.85}

\usepackage{graphicx}

\begin{document}

\title{Distributed Blind Source Separation based on F\MakeLowercase{ast}ICA}

\author{Cem Ates Musluoglu, Alexander Bertrand, \IEEEmembership{Senior Member, IEEE}
\thanks{Copyright \copyright 2025 IEEE. Personal use of this material is permitted. Permission from IEEE must be obtained for all other uses, in any current or future media, including reprinting/republishing this material for advertising or promotional purposes, creating new collective works, for resale or redistribution to servers or lists, or reuse of any copyrighted component of this work in other works.}
\thanks{This project has received funding from the European Research Council (ERC) under the European Union's Horizon 2020 research and innovation programme (grant agreement No 802895 and No 101138304). The authors also acknowledge the financial support of the FWO (Research Foundation Flanders) for project G081722N, and the Flemish Government (AI Research Program). Views and opinions expressed are however those of the author(s) only and do not necessarily reflect those of the European Union or ERC. Neither the European Union nor the granting authority can be held responsible for them.}
\thanks{C. A. Musluoglu and A. Bertrand are with KU Leuven, Department of Electrical Engineering (ESAT), Stadius Center for Dynamical Systems, Signal Processing and Data Analytics, Kasteelpark Arenberg 10, box 2446, 3001 Leuven, Belgium and with Leuven.AI - KU Leuven institute for AI. e-mail: cemates.musluoglu, alexander.bertrand @esat.kuleuven.be}}

\maketitle

\begin{abstract}
   With the emergence of wireless sensor networks (WSNs), many traditional signal processing tasks are required to be computed in a distributed fashion, without transmissions of the raw data to a centralized processing unit, due to the limited energy and bandwidth resources available to the sensors. In this paper, we propose a distributed independent component analysis (ICA) algorithm, which aims at identifying the original signal sources based on observations of their mixtures measured at various sensor nodes. One of the most commonly used ICA algorithms is known as FastICA, which requires a spatial pre-whitening operation in the first step of the algorithm. Such a pre-whitening across all nodes of a WSN is impossible in a bandwidth-constrained distributed setting as it requires to correlate each channel with each other channel in the WSN. We show that an explicit network-wide pre-whitening step can be circumvented by leveraging the properties of the so-called Distributed Adaptive Signal Fusion (DASF) framework. Despite the lack of such a network-wide pre-whitening, we can still obtain the $Q$ least Gaussian independent components of the centralized ICA solution, where $Q$ scales linearly with the required communication load.
\end{abstract}

\begin{IEEEkeywords}
   Distributed Optimization, Distributed Spatial Filtering, Independent Component Analysis.
\end{IEEEkeywords}

\IEEEpeerreviewmaketitle

\section{Introduction}
\label{sec:intro}

\IEEEPARstart{I}{ndependent} component analysis (ICA) is a well-known method for reconstructing statistically independent source signals from linear mixtures of these sources measured across multiple sensors \cite{hyvarinen1999fast,hyvarinen2000independent,hyvarinen2013independent}. It has generally been associated with the blind source separation problem \cite{choi2005blind} and has found applications in various fields including biomedical \cite{james2004independent,zhukov2000independent} or acoustic signal processing \cite{asano2003combined} among others \cite{naik2011overview,bouveresse2016independent}. The emergence and versatility of wireless sensor networks (WSNs) \cite{akyildiz2002wireless,kumar2019machine} make it attractive to solve such problems in a distributed setting, where different channels of the mixture signal are measured across various sensors placed in different locations. However, the energy and bandwidth bottlenecks of WSNs typically limit the amount of data that can be transmitted between the nodes and therefore make the use of a fusion center often not feasible in practice. Therefore, distributed algorithms that avoid data centralization are necessary for solving the ICA problem in a distributed setting such as WSNs.

Distributed algorithms have been proposed to solve the ICA / blind source separation problem \cite{hioka2011distributed,alavi2016distributed}. In \cite{hioka2011distributed}, the network-wide ICA problem is solved heuristically with a (suboptimal) divide and conquer-type approach based on local neighborhoods. In \cite{alavi2016distributed}, the ICA problem is solved in a truly distributed fashion, using the well-known alternative direction method of multipliers (ADMM). However, this algorithm iterates over sample batches, requiring multiple retransmissions of (updated versions of) the same batch of samples at each node, which leads to a large communication cost scaling poorly with the network size, often sharing even more data than what was collected at the node. Furthermore, for each new incoming batch of samples, the ADMM iterations have to start from scratch, making ADMM-based algorithms unfit for (adaptive) spatial filtering on streaming data \cite{musluoglu2022unifiedp1}. We note that most distributed estimation methods based on ADMM, consensus, or diffusion \cite{olfati2005consensus,cattivelli2009diffusion,zayyani2021robust,ciuonzo2021distributed} were originally designed for a setting where the data is partitioned in the sample dimension, making them unfit for the distributed ICA problem, where the data is partitioned in the channel dimension. Although some of these methods can be adapted to a channel-partitioning setting, they would face similar challenges as the ADMM-based ICA method in  \cite{alavi2016distributed}, requiring multiple iterations over every new sample batch. Another potential approach would be to first perform a distributed dimensionality reduction algorithm and perform a centralized ICA on the projected data. However, this is not always applicable and might lead to suboptimal results. For example, using a distributed principal component analysis procedure \cite{bertrand2014dacmee}, we would select the subspace of sources with the highest variance when reducing dimensions. In low SNR scenarios, this high-variance subspace could mainly capture noise sources, instead of the relevant underlying sources \cite{artoni2018applying}.

In this paper, we propose DistrICA, a distributed algorithm that directly solves the network-wide ICA problem in an adaptive setting with streaming data, without first projecting the sensor data in a low-dimensional subspace. By only sharing linearly fused sensor signals between the nodes, the communication burden is significantly reduced, while still obtaining a subset of the sources from the centralized ICA problem.

\section{Problem Setting}
\label{sec:prob_setting}

Let us consider $M$ sensor signals $y_m$, $m\in\{1,\dots,M\}$, measured at each sensor $m$ and denote by $y_m(t)\in\mathbb{R}$ the observation of $y_m$ at time $t$. Arranging these signals as $\mathbf{y}(t)=[y_1(t), \dots, y_M(t)]^T\in\mathbb{R}^M$, we make the assumption that
\begin{equation}\label{eq:signal_model}
  \mathbf{y}(t)=A\cdot \mathbf{s}(t),
\end{equation}
where $A\in\mathbb{R}^{M\times M}$ is the mixture matrix, and $\mathbf{s}(t)\in\mathbb{R}^{M}$ is the vector containing statistically independent source signals $s_m(t)$, $m\in\{1,\dots,M\}$. In the remaining parts of this paper, we will omit the time index $t$ unless we want to explicitly refer to a specific time sample. Our objective is to recover the sources $s_m$ from observations of $\mathbf{y}$ by applying a linear filter $X^*$ such that $X^{*T}\mathbf{y}=\mathbf{r}$, where the entries of $\mathbf{r}$ are equal to those of $\mathbf{s}$ up to a scaling and permutation ambiguity \cite{hyvarinen2000independent}. Note that this can also be a partial recovery, in the sense that we might be only interested in a subset of the sources $s_m$, i.e., $\mathbf{r}\in\mathbb{R}^Q$, with $Q\leq M$. There exists various methods for this purpose, such as maximizing non-Gaussianity, maximum likelihood estimation, or minimizing the mutual information \cite{hyvarinen2000independent,oja2006fastica}. In this paper, we focus on the FastICA algorithm \cite{hyvarinen1999fast}, which consists of first applying a pre-whitening step on $\mathbf{y}$ to obtain a signal $\mathbf{z}$, followed by an orthogonal transformation that maximizes the ``non-Gaussianity'' of the demixed signals\footnote{The core idea behind this is that a mixture of non-Gaussian sources is closer to a Gaussian distribution than any of the unmixed sources.}. Formally, the whitening procedure can be written as
\begin{equation}\label{eq:whitening}
  \mathbf{z}=ED^{-1/2}E^T\mathbf{y},
\end{equation}
such that $\mathbb{E}[\mathbf{z}\mathbf{z}^T]=I$, where $E$ and $D$ are obtained from the eigenvalue decomposition of the covariance matrix of $\mathbf{y}$, i.e., $R_{\mathbf{yy}}=\mathbb{E}[\mathbf{y}\mathbf{y}^T]=EDE^T$, assumed to be full rank\footnote{This is to guarantee well-posedness \cite{hadamard1902problemes,zhou2005hadamard} of the problem, which can be briefly summarized as requiring that a small change in the inputs of the problem (the data) implies a small change in the output (the optimal solution).}. The ICA filters are then obtained by solving the following problem
\begin{equation}\label{eq:ica_prob_w}
  \underset{W=[\mathbf{w}_1,\dots,\mathbf{w}_Q]}{\text{max. } } \quad  \sum_{m=1}^Q\mathbb{E}[F(\mathbf{w}_m^T\mathbf{z})] \quad \textrm{s.t. } \; W^TW=I.
\end{equation}
$F$ is often chosen as $F(x)=\text{log cosh}(x)$ or $F(x)=-\text{exp}(-x^2/2)$, making $\mathbb{E}[F(x)]$ a proxy for the negentropy of the random variable $x$ (also known as the non-Gaussianity) \cite{hyvarinen2000independent}. The $Q$ least Gaussian independent components are then found by applying a solution $W^*$ of (\ref{eq:ica_prob_w}) to $\mathbf{z}:W^{*T}\mathbf{z}$. Equivalently, we can express this result using the original data $\mathbf{y}$ as $X^{*T}\mathbf{y}$, where $X^*=ED^{-1/2}E^TW^*$ from (\ref{eq:whitening}). The steps of FastICA are presented in Algorithm \ref{alg:fastica}, where $F'$ and $F''$ correspond to the first and second order derivative of $F$.

Let us now consider a sensor network represented by a graph $\mathcal{G}$ with $K$ nodes within the set $\mathcal{K}=\{1,\dots,K\}$. Each node $k$ measures an $M_k-$channel signal $\mathbf{y}_k$, such that, defining
\begin{equation}\label{eq:y_part}
  \mathbf{y}=[\mathbf{y}_1^T,\dots,\mathbf{y}_K^T]^T,
\end{equation}
our goal is to solve the ICA problem on the network-wide data $\mathbf{y}\in\mathbb{R}^M$, where $M=\sum_k M_k$. The signal $\mathbf{y}$ is assumed to be ergodic and (short-term) stationary, such that its statistical properties can be estimated through temporal averaging of observed samples. In such a distributed setting, our aim is again to find a spatial filter $X^*\in\mathbb{R}^{M\times Q}$ such that $X^{*T}\mathbf{y}$ provides the $Q$ least Gaussian sources in $\mathbf{s}$. However, the pre-whitening (\ref{eq:whitening}) of the data requires the knowledge of the spatial correlation between different channels of $\mathbf{y}$. While there exist distributed algorithms for obtaining eigenvectors of $R_{\mathbf{yy}}$ (see, e.g., \cite{bertrand2014dacmee,li2011distributed}), these methods can only extract a few principal eigenvectors (depending on the available communication budget), and therefore can only compute a compressive version of (\ref{eq:whitening}) in which the data is projected onto a lower-dimensional subspace, which is not always desirable \cite{artoni2018applying}.

\begin{figure}[!t]
  \removelatexerror
  \begin{algorithm}[H]
  \caption{FastICA \cite{hyvarinen1999fast}}\label{alg:fastica}
  \DontPrintSemicolon
  \SetKwInOut{Input}{input}\SetKwInOut{Output}{output}
  \BlankLine
  \Input{Multi-channel signal $\mathbf{y}$}
  $i\gets0$, $W\gets [\;]$\;
  Compute $EDE^T$ as the eigenvalue decomposition of $R_{\mathbf{yy}}$\;
  $\mathbf{z}\gets ED^{-1/2}E^T\mathbf{y}$\;
    \For{$m\in \{1,\dots,Q\}$}{
      $\mathbf{w}$ initialized randomly\;
      \While{convergence criterion not reached}
      {
      $\mathbf{w}\gets \mathbb{E}[\mathbf{z}F'(\mathbf{w}^T\mathbf{z})]-\mathbb{E}[F''(\mathbf{w}^T\mathbf{z})]\mathbf{w}$\;
      $\mathbf{w}\gets \mathbf{w}-WW^T\mathbf{w}$ if $m>1$\;
      $\mathbf{w}\gets \mathbf{w}/\|\mathbf{w}\|$\;
      }
      $W\gets [W,\mathbf{w}]$\;
    }
    $X\gets ED^{-1/2}E^TW$
  \end{algorithm}
\end{figure}

\section{The Distributed ICA algorithm}
\label{sec:districa}
In this section, we present a distributed ICA method based on the Distributed Adaptive Signal Fusion (DASF) framework \cite{musluoglu2022unifiedp1,musluoglu2022unifiedp2,hovine2024distributed,hovine2023distributed,musluoglu2023dansf}, which allows solving (adaptive) channel-partitioned distributed optimization problems. DASF uses a data-driven fuse-and-forward approach, where the $N$ most recent samples of all nodes are forwarded towards an updating node, while linearly fusing them along the way. The updating node collects these fused streams and locally solves a compressed version of the original network-wide problem to find the in-network fusion rules for the next iteration. By rotating the updating node at each iteration, convergence to the optimal solution can be achieved \cite{musluoglu2022unifiedp2,hovine2024distributed}. However, the ICA problem does not trivially fit the family of problems that the DASF algorithm can solve. Nevertheless, in Section \ref{sec:reformulation}, we will explain how the two-step problem (\ref{eq:whitening})-(\ref{eq:ica_prob_w}) can be cast into the DASF framework, and in particular, how the across-node pre-whitening step (\ref{eq:whitening}) can be bypassed, such that only per-node pre-whitening operations are required.
\subsection{Reformulating ICA as a DASF problem} \label{sec:reformulation}
As described in \cite{musluoglu2022unifiedp1}, the general form that the problems fitting the DASF framework take is given by
\begin{equation}\label{eq:dasf_prob}
  \begin{aligned}
  \underset{X}{\text{min. } } \mathbb{E}[G(X^T\mathbf{y})]\quad
  \textrm{s.t. } & \mathbb{E}[H_j(X^T\mathbf{y})]\leq 0\;\textrm{ $\forall j\in\mathcal{J}_I$,}\\
    & \mathbb{E}[H_j(X^T\mathbf{y})]=0\;\textrm{ $\forall j\in\mathcal{J}_E$,}
  \end{aligned}
\end{equation}
where the $H_j$'s denote constraint functions of the problem and the sets $\mathcal{J}_I$ and $\mathcal{J}_E$ correspond to index sets of inequality and equality constraints respectively. The DASF algorithm can solve problems of the form (\ref{eq:dasf_prob}) in a fully distributed and time-adaptive fashion with provable convergence guarantees \cite{musluoglu2022unifiedp2}. Note that the family of the problems fitting the DASF framework is larger than the one represented by the formulation in (\ref{eq:dasf_prob}), which is omitted here for conciseness. Let us now cast (\ref{eq:whitening})-(\ref{eq:ica_prob_w}) into a problem fitting the DASF framework by embedding the pre-whitening step (\ref{eq:whitening}) within the ICA optimization problem (\ref{eq:ica_prob_w}), by making the change of variables: $X=ED^{-1/2}E^TW$, resulting in the equivalent problem
\begin{equation}\label{eq:ica_prob}
  \underset{X=[\mathbf{x}_1,\dots,\mathbf{x}_Q]}{\text{max. } } \quad \sum_{m=1}^Q\mathbb{E}[F(\mathbf{x}_m^T\mathbf{y})] \quad\textrm{s.t. } \; X^TR_{\mathbf{yy}}X=I.
\end{equation}
Then, we can rewrite the objective function of (\ref{eq:ica_prob}) as $\mathbb{E}[G(X^T\mathbf{y})]$, where $G(X^T\mathbf{y})=\sum_{m=1}^QF_m(X^T\mathbf{y})$, such that $F_m(X^T\mathbf{y})=F(\mathbf{e}_m^TX^T\mathbf{y})$, where $\mathbf{e}_m$ is the $m-$th column of the $Q\times Q$ identity matrix, making it fit the formulation in (\ref{eq:dasf_prob}). Additionally, note that each entry of the constraint $X^TR_{\mathbf{yy}}X=\mathbb{E}[X^T\mathbf{yy}^TX]=I$ is written as $\mathbb{E}[\mathbf{x}_m^T\mathbf{yy}^T\mathbf{x}_n-\mathds{1}\{m=n\}]=\mathbb{E}[\mathbf{e}_m^TX^T\mathbf{yy}^TX\mathbf{e}_n-\mathds{1}\{m=n\}]=0$, for $m,n\in\{1,\dots Q\}$, where $\mathds{1}$ denotes the indicator function. Therefore, taking
\begin{equation}
  H_{m,n}(X^T\mathbf{y})=\mathbf{e}_m^TX^T\mathbf{yy}^TX\mathbf{e}_n-\mathds{1}\{m=n\},
\end{equation}
we see that the constraints of (\ref{eq:ica_prob}) fit the formulation of the DASF framework in (\ref{eq:dasf_prob}). Note that, since (\ref{eq:ica_prob}) is equivalent to (\ref{eq:whitening})-(\ref{eq:ica_prob_w}), it should be clear that Algorithm \ref{alg:fastica} also defines a solver for (\ref{eq:ica_prob}). The latter is a seemingly trivial yet important observation, which we will exploit in the next subsections when deriving the proposed DistrICA algorithm.

\subsection{Data flow of DistrICA}

Within the distributed problem setting introduced in Section \ref{sec:prob_setting}, let every node $k\in\mathcal{K}$ have an estimate $X_k^i$ at iteration $i$ of the block $X_k\in\mathbb{R}^{M_k\times Q}$ of $X$, partitioned as $\mathbf{y}$ in (\ref{eq:y_part}):
\begin{equation}
  X=[X_1^T,\dots,X_K^T]^T.
\end{equation}
All $X_k^i$'s are initialized randomly if $i=0$. Every node then collects $N$ time samples of its own $M_k-$channel signal $\mathbf{y}_k$ to obtain $\{\mathbf{y}_k(t)\}_{t=iN}^{iN+N-1}$ and linearly compresses every sample using its current estimate\footnote{Note that the amount of compression depends on $Q$, which is defined by the number of independent components that we wish to extract.} of $X_k$ into the $Q-$channel signal
\begin{equation}\label{eq:yhat}
  \widehat{\mathbf{y}}_k^i(t)=X_k^{iT}\mathbf{y}_k(t).
\end{equation}
Let us select one node among all the nodes of the network to be the updating node for the current iteration $i$, which we call $q$. The network is then temporarily pruned into a tree $\mathcal{T}^i(\mathcal{G},q)$ to obtain a unique path between each and every node. The pruning function should not remove any link between the updating node $q$ and its neighbors, but can otherwise be chosen freely \cite{musluoglu2022unifiedp1}. Every node then proceeds to transmit the $N$ compressed samples $\{\widehat{\mathbf{y}}_k^i(t)\}_{t=iN}^{iN+N-1}$, towards this updating node $q$ in an inwards flow in which signals from neighboring nodes are linearly fused throughout their path towards node $q$:
\begin{equation}\label{eq:sum_fwd}
  \widehat{\mathbf{y}}_{k \rightarrow n}^i(t)\triangleq X_k^{iT}\mathbf{y}_k(t)+\sum_{l\in\mathcal{N}_k\backslash\{n\}}\widehat{\mathbf{y}}_{l\rightarrow k}^i(t),
\end{equation}
where $\mathcal{N}_k$ denotes the set of neighboring nodes of node $k$ after pruning, and $\widehat{\mathbf{y}}_{k\rightarrow n}^i\in\mathbb{R}^{Q}$ is the data transmitted by node $k$ to its neighboring node $n$. Note that the second term of (\ref{eq:sum_fwd}) is recursive and vanishes for the leaf nodes of the tree, which initializes the process such that the inward flow naturally arises without central coordination (a node $l$ sends its data to node $n$ from the moment it has received data from all its neighbors except one, being node $n$). In this way, each node effectively transmits a $Q-$channel signal, independent of the size of the network, making the communication bandwidth fully scalable.

The updating node $q$ eventually receives $N$ samples of
\begin{equation}\label{eq:sum_fwd_n}
  \widehat{\mathbf{y}}_{n\rightarrow q}^i(t)=X_n^{iT}\mathbf{y}_n(t)+\sum_{k\in\mathcal{N}_n\backslash\{q\}}\widehat{\mathbf{y}}_{k\rightarrow n}^i(t)=\sum_{k\in\mathcal{B}_{nq}}\widehat{\mathbf{y}}_k^i(t)
\end{equation}
from all its neighbors $n\in\mathcal{N}_q$, where $\mathcal{B}_{nq}$ is the branch of $\mathcal{T}^i(\mathcal{G},q)$ that is connected to $q$ via $n$ (excluding $q$ itself).

\subsection{Updating step}

With the data exchange described in the previous subsection, node $q$ is now in possession of $N$ samples of $\widehat{\mathbf{y}}_{n\rightarrow q}^i$ for $n\in\mathcal{N}_q$ and $N$ samples of its own (uncompressed) signal $\mathbf{y}_q$. Stacking these, we define the data available at node $q$ and iteration $i$ as
\begin{equation}\label{eq:y_tilde}
  \widetilde{\mathbf{y}}_q^i(t)\triangleq[\mathbf{y}_q^T(t),\widehat{\mathbf{y}}_{n_1\rightarrow q}^{iT}(t),\dots,\widehat{\mathbf{y}}_{n_{|\mathcal{N}_q|}\rightarrow q}^{iT}(t)]^T\;\in\mathbb{R}^{\widetilde{M}_q},
\end{equation}
where $\widetilde{M}_q=|\mathcal{N}_q|\cdot Q+M_q$. At this step, the DASF algorithm \cite{musluoglu2022unifiedp1} requires the updating node $q$ to use the available signal, i.e., $\widetilde{\mathbf{y}}_q^i$ to solve the following compressed version of (\ref{eq:ica_prob}):
\begin{equation}\label{eq:ica_prob_compressed}
    \underset{\widetilde{X}_q=[\widetilde{\mathbf{x}}_1,\dots,\widetilde{\mathbf{x}}_Q]}{\text{max. } } \quad \sum_{m=1}^Q\mathbb{E}[F(\widetilde{\mathbf{x}}_m^T\widetilde{\mathbf{y}}_q^i)]\quad 
    \textrm{s.t. } \; \widetilde{X}_q^TR^i_{\widetilde{\mathbf{y}}_q\widetilde{\mathbf{y}}_q}\widetilde{X}_q=I,
\end{equation}
where $R^i_{\widetilde{\mathbf{y}}_q\widetilde{\mathbf{y}}_q}=\mathbb{E}[\widetilde{\mathbf{y}}_q^i\widetilde{\mathbf{y}}_q^{iT}]$ is the covariance matrix of $\widetilde{\mathbf{y}}_q^i$. As seen in (\ref{eq:ica_prob_compressed}), a new variable $\widetilde{X}_q$ is defined to act analogously to $X$ locally. Remarkably, (\ref{eq:ica_prob_compressed}) has the exact same form as (\ref{eq:ica_prob}), and can therefore be solved locally at node $q$ by applying the FastICA algorithm described in Algorithm \ref{alg:fastica}, based on the batch of $N$ available samples of $\widetilde{\mathbf{y}}_q^i$. We have thus effectively bypassed the network-wide pre-whitening step (\ref{eq:whitening}) and replaced it with a local one instead at node $q$.

While the solution $\widetilde{X}_q^*$ of (\ref{eq:ica_prob_compressed}) is only defined up to a scaling and permutation ambiguity, the solver in Algorithm \ref{alg:fastica} always finds the solution where the sources are scaled to unit-norm and ordered from least to most Gaussian. The remaining sign ambiguity can be easily resolved, e.g., by ensuring that the largest value of each column of $\widetilde{X}_q^*$ is positive, in order to avoid spurious sign flips across iterations.

\begin{figure}[!t]
  \removelatexerror
  \DontPrintSemicolon
  \begin{algorithm}[H]
  \caption{DistrICA Algorithm}\label{alg:DistrICA}
  \SetKwInOut{Output}{output}
  \BlankLine
  $X^0$ initialized randomly, $i\gets0$.\;
  \Repeat
  {
  Choose the updating node as $q\gets (i\mod K)+1$.\;
  1) The network $\mathcal{G}$ is pruned into a tree $\mathcal{T}^i(\mathcal{G},q)$.\;

  2) All nodes $k$ collect $N$ samples of $\mathbf{y}_k$ and compress them into $\widehat{\mathbf{y}}^{i}_k$.\;

  3) The nodes sum-and-forward $\widehat{\mathbf{y}}^{i}_k$ towards node $q$ via the recursive rule (\ref{eq:sum_fwd}). Node $q$ eventually receives $N$ samples of $\widehat{\mathbf{y}}^{i}_{n\rightarrow q}$ given in (\ref{eq:sum_fwd_n}), from all its neighbors $n\in\mathcal{N}_q$.\; 

  \At{Node $q$}
  {
    4a) Solve Problem (\ref{eq:ica_prob_compressed}) to obtain $\widetilde{X}_q^*$ using Algorithm \ref{alg:fastica} on $\widetilde{\mathbf{y}}_q^i$ defined in (\ref{eq:y_tilde}).\;
    4b) Extract the $Q$ least Gaussian sources by computing $\widetilde{X}_q^{*T}\widetilde{\mathbf{y}}^i_q$.\;
    4c) Partition $\widetilde{X}^{*}_q$ as in (\ref{eq:X_tilde_part}) and disseminate every $G_n^{i+1}$ in the corresponding subgraph $\mathcal{B}_{nq}$.\;
  }
  
  5) Every node updates $X_k^{i+1}$ according to (\ref{eq:upd_tree}).\;
  
  $i\gets i+1$\;
  }
  \end{algorithm}
\end{figure}

The solution $\widetilde{X}_q^*$ of (\ref{eq:ica_prob_compressed}) is then partitioned as
\begin{equation}\label{eq:X_tilde_part}
  \widetilde{X}_q^{*}=[X_q^{(i+1)T},G_{n_1}^{(i+1)T},\dots,G_{n_{|\mathcal{N}_q|}}^{(i+1)T}]^T,
\end{equation}
such that each partition has a corresponding block in the partitioning of $\widetilde{\mathbf{y}}_q^i$ as defined in (\ref{eq:y_tilde}). The $Q$ least Gaussian sources can then be extracted at node $q$ by computing $\widetilde{X}_q^{*T}\widetilde{\mathbf{y}}^i_q=X^{(i+1)T}\mathbf{y}$, and be transmitted from node $q$ to other nodes acting as data sinks if necessary. The blocks $G_{n}^{i+1}$ are disseminated into the corresponding subgraph $\mathcal{B}_{nq}$ through node $n$, allowing every node to update its local estimator as
\begin{equation}\label{eq:upd_tree}
  X_k^{i+1}=\begin{cases}
  X_q^{i+1} & \text{if $k=q$} \\
  X_k^{i}G_n^{i+1} & \text{if $k\in\mathcal{B}_{nq}$, $n\in\mathcal{N}_q$},
  \end{cases}
\end{equation}
such that, at the end of iteration $i$, the new estimate of the network-wide variable $X$ becomes
\begin{equation}
  X^{i+1}=[X_1^{(i+1)T},\dots,X_K^{(i+1)T}]^T.
\end{equation}
This procedure is then repeated with a new updating node (e.g., chosen in a round-robin fashion) and a new set of $N$ samples of $\mathbf{y}$ at each iteration, such that changes in the statistical properties can be tracked, making the algorithm adaptive. In particular, $X^i$ is an estimator of $X^*(t)$ for $t=iN$. The steps of the proposed DistrICA algorithm are summarized in Algorithm \ref{alg:DistrICA}. At the updating node, the computational complexity of DistrICA is in the order of the one of FastICA applied on the compressed data $\widetilde{\mathbf{y}}_q$, namely $\mathcal{O}(\widetilde{M}_q^2(N+\widetilde{M}_q)+QT(\widetilde{M}_qN+\widetilde{M}_q^2Q))$, where $T$ corresponds to the number of iterations of the FastICA loop.

Since DistrICA is derived according to the technical principles of the DASF framework, its convergence is guaranteed by the theoretical results in \cite{musluoglu2022unifiedp2}. Note that \cite{musluoglu2022unifiedp2} imposes some mild -- yet highly technical -- conditions, which can be shown to hold for the case of DistrICA, but which are omitted here for conciseness. One crucial condition is that the number of constraints in (\ref{eq:dasf_prob}) has to be smaller than $Q^2$, which is satisfied since there are $Q(Q+1)/2$ constraints in (\ref{eq:ica_prob}) due to the symmetry of $R_{\mathbf{yy}}$. Additionally, we exploit the fact that the solution set of FastICA is finite, where solutions can only differ up to a sign change of their columns, instead of an invariance to scaling and permutations, guaranteeing convergence to a single point \cite[Theorem 5]{musluoglu2022unifiedp2}. As (\ref{eq:ica_prob}) is a non-convex problem, the results in \cite{musluoglu2022unifiedp2} only guarantee convergence to a stationary point of (\ref{eq:ica_prob_w}), which is also the case for the centralized FastICA algorithm. Nevertheless, FastICA almost always converges to a global optimum in practice \cite{oja2006fastica}, such that the same holds for DistrICA.

\begin{figure}[t]
  \captionsetup[subfigure]{width=.75\textwidth}
  \centering
  \begin{subfigure}[b]{0.48\textwidth}
      \centering
      \includegraphics[width=\textwidth]{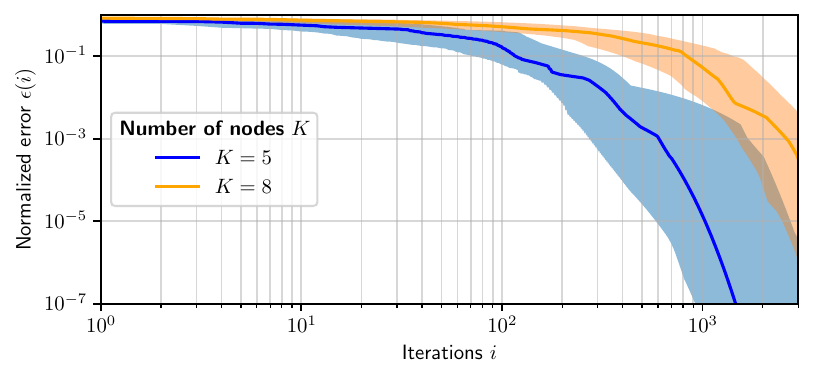}
  \end{subfigure}
  \begin{subfigure}[b]{0.48\textwidth}
      \centering
      \includegraphics[width=\textwidth]{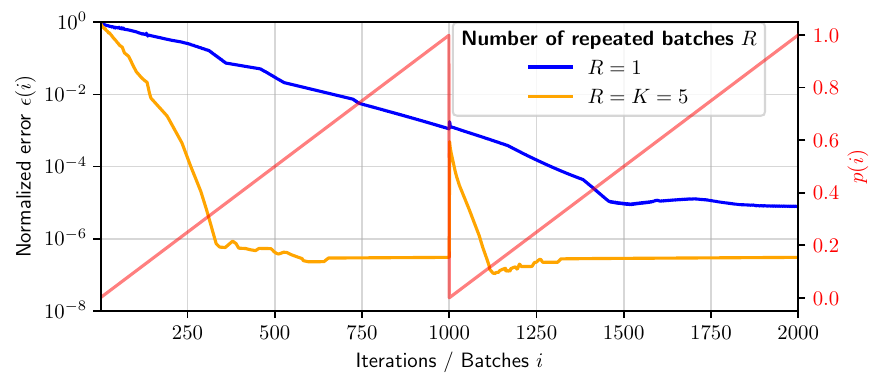}
  \end{subfigure}
  \begin{subfigure}[b]{0.48\textwidth}
      \centering
      \includegraphics[width=\textwidth]{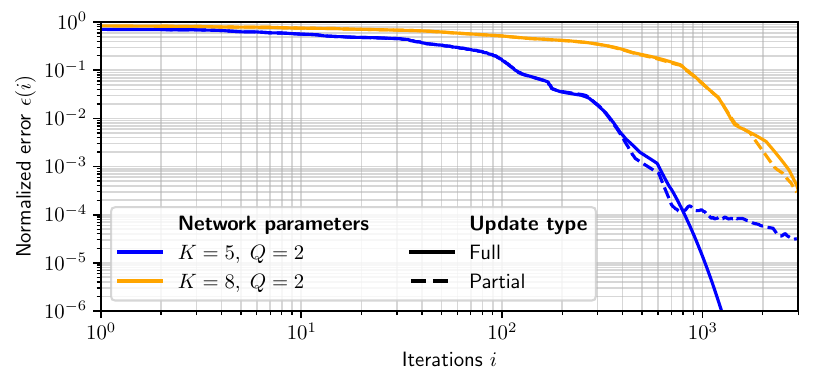}
  \end{subfigure}
  \caption{(Top) Normalized error for varying number of nodes $K$. (Middle) Convergence in an adaptive setting. (Bottom) Comparison of the error for DistrICA with full and partial solutions.}\label{fig:nbnodes_nbfilters}
\end{figure}

\section{Simulations}
\label{sec:sim}

We consider a wireless sensor network where each node has $M_k=5$ sensors, and therefore measure a $5-$channel local signal $\mathbf{y}_k$. Throughout this section, we consider networks randomly generated using the Erd\H{o}s-R\'{e}nyi model with connection probability equal to $0.8$ and take $F(x)=\text{log cosh}(x)$ in the objective of (\ref{eq:ica_prob}). The signal model is as given in (\ref{eq:signal_model}) and generated as follows. The elements of the mixture matrix $A\in\mathbb{R}^{M\times M}$ are drawn independently at random from the standard Gaussian distribution, i.e., $\mathcal{N}(0,1)$. We consider $M$ independent sources in $\mathbf{s}$, where $s_1$ is a sinusoid, and $s_2$ a square signal (i.e., $Q=2$), while $s_m$, $2<m\leq M$ are convex combinations of uniformly and normally distributed noise, i.e., $s_m(t)=\alpha_m u_m(t)+(1-\alpha_m)n_m(t)$, where $u_m\sim \mathcal{U}[-0.5,0.5]$, $n_m\sim \mathcal{N}(0,1)$, and $\alpha_m\in[0,1]$ is different for each source $m$. The resulting mixtures observed at the different sensors are normalized to unit variance. The goal is to separate $s_1$ and $s_2$ from the near-Gaussian (noise) sources. Note that all sources $\mathbf{s}$ have more or less the same variance, hence a principal component analysis will not be able to separate the target subspace from the noise subspace. 

To assess the convergence of DistrICA to the centralized solution, we first consider a stationary setting, where at each iteration, $N=10^4$ samples are measured at the sensor nodes, and the necessary statistical parameters of the stochastic signals are approximated using a temporal averaging over these samples. We use the normalized squared error to measure the accuracy of the estimator $\{X^i\}_{i}$ obtained through DistrICA over iterations $i$, given by $\varepsilon(i)=\text{median}\left(\frac{\|X^i-X^*\|^2}{\|X^*\|^2}\right)$, where the median is taken over $30$ Monte-Carlo runs under the same experimental settings. An optimal solution $X^*$ of the centralized ICA problem (\ref{eq:ica_prob}) is used for comparison and obtained through FastICA in a centralized setting. All simulations have been performed using the DASF toolbox \cite{musluoglu2022dsfotoolbox}.

Fig. \ref{fig:nbnodes_nbfilters} (Top) shows convergence plots for two different network sizes: $K=5$ and $K=8$. We observe that, although both settings lead to convergence towards an optimal solution of ICA, the setting with fewer nodes does so faster. This is expected, as the per-node updating frequency is lower in a larger network due to the sequential updating procedure. 

In Fig. \ref{fig:nbnodes_nbfilters} (Middle), we show the convergence of DistrICA in an adaptive setting, where the mixture matrix $A$ changes in time. For each new incoming batch of samples, $A$ is updated to $A+\Delta\cdot p(i)$, where $p$ is a scalar function defined by the red curve in Fig. \ref{fig:nbnodes_nbfilters}, changing at each iteration $i$, i.e., every batch of $N$ samples will have a different mixture matrix. The entries of $\Delta$ are drawn from $\mathcal{N}(0,1)$ first, and then scaled such that $\|\Delta\|_F=0.005\cdot\|A\|_F$. To make the DistrICA algorithm more adaptive, we now also implement the alternative strategy from \cite{musluoglu2022improved}, which reuses sample batches $R$ times over multiple iterations (see \cite{musluoglu2022improved} for details, $R=1$ corresponds to the original algorithm). Note that, in contrast to the stationary setting, we now observe a tracking error that saturates at a non-zero value due to changing signal statistics.

Fig. \ref{fig:nbnodes_nbfilters} (Bottom) presents results of DistrICA when the FastICA solver of the compressed problem (\ref{eq:ica_prob_compressed}) is not exact, i.e., where we only let the nodes partially solve their local compressed ICA problem. In this experiment, we stop the local FastICA algorithm when it reaches an error of $10^{-3}$ in the normed difference between two consecutive filter values or a maximum of $10$ iterations. This allows reducing the computational burden at the nodes while still guaranteeing convergence \cite{hovine2024distributed}. We see that until a certain error level, the convergence is not affected by this partial updating scheme and follows similar convergence properties while greatly reducing the number of computations performed at each node.

\section{Conclusion}

We have proposed the DistrICA algorithm to solve the ICA problem in a distributed and adaptive setting without requiring centralization of the data measured at different sensor nodes. After presenting its technical aspects, we have provided extensive simulation results comparing different problem settings and validating its convergence towards the centralized solution of the ICA problem.

\bibliographystyle{IEEEbib}
\bibliography{refs}

\end{document}